\documentstyle{sup}
\input {psfig.tex}
\vspace{-8pt}

\title[Superlattices and Microstructures, Vol.\ 20, No.\ 1, 1996]
{A microscopic theory of skyrmion excitations in the\\
fractional quantum Hall system}
\author[Superlattices and Microstructures, Vol.\ 20, No.\ 1, 1996]
{Kang-Hun Ahn and K. J. Chang%
\cr\vspace{10pt}%
{\normalsize\it Department of Physics, Korea Advanced Institute 
of Science and Technology, Taejon 305-701, Korea}\cr
}

\pagerange{\pageref{firstpage}--\pageref{lastpage}}

\def\LaTeX{L\kern-.25em\raise.425ex\hbox{a}\kern-.075em\TeX}

\def\fakebold#1{\relax\ifvmode\leavevmode\fi%
\ifmmode%
\setbox0=\hbox{$#1$}%
\else%
\setbox0=\hbox{#1}%
\fi%
\kern-.02em\copy0 \kern-\wd0%
\kern .04em\copy0 \kern-\wd0%
\kern-.0125em\raise.02em\box0%
}%

\begin{document}
\label{firstpage}
\maketitle
\sloppy
\begin{center}
\received{(Received 15 July 1996)}
\end{center}
%
\begin{abstract}
We present a microscopic theory of skyrmion and antiskyrmion excitations
in fractional quantum Hall systems, and calculate in an analytical fashion
their excitation energies.
From the calculated net spins at various fractional filling factors,
we find the magnetic field dependence of the spin polarization of
the skyrmion-condensated state.
\end{abstract}

\section{Introduction}

Charged spin textures known as skyrmions in the quantum Hall system
have been of great interest.
In the limit of weak Zeeman coupling, it was predicted that the lowest
energy excitations around filling factor $\nu$ = 1/$M$ with odd integer
$M$ are skyrmions \cite{sondhi}.
Recent experiments demonstrated the existence of skyrmions for $\nu$
near 1 in two-dimensional electron systems \cite{barret,schmeller,aifer}.
Since the skyrmion size is determined by relative strength of Coulomb
and Zeeman interaction, the effective field theory of nonlinear sigma
(NL$\sigma$) model has not been successful in describing correctly the
observed spin polarization \cite{barret,schmeller,aifer}.
Despite several microscopic theoretical attempts to overcome the drawback
of the field theory \cite{fertig,nayak,moon}, most calculations
were limited to the case of $\nu=1$.

In this paper, we develop a microscopic approach to the skyrmion in the
fractional quantum Hall regime, which allows for analytical calculations
of the total energies and net spins of skyrmion and antiskyrmion.
We perform infinite-size calculations for filling factors $\nu$ = 1, 4/5,
2/3, 1/3, and 1/5, and predict the field dependence of the spin
polarization for the skyrmion-condensated state.

\section{Theory}

At filling factor $\nu$ = 1/$M$, the ground state of a quantum Hall
system is well approximated by the incompressible spin-polarized
state in the symmetric gauge:
\begin{eqnarray}
\Phi_{g}^{\nu} = \sum_{i_{1} < i_{2} < ... < i_{N} }
C^{\nu} (i_{1} , i_{2} , ..., i_{N} ) \prod_{k=1}^{N}
c^{ \dagger }_{i_k,\uparrow} \mid 0 \rangle, 
\label{ground}
\end{eqnarray}
where $\mid 0 \rangle$ is the vacuum state and $c_{m,s}^{\dagger}$ creates
an electron of spin $s$ and angular momentum $m$ in the lowest Landau level.
Here the coefficients $C^{\nu}$ can be determined by expanding the known
Laughlin wave function.
In analogy, the variational wave function for a  charged
spin-texture excitation will be of the form 
\begin{eqnarray}
\Psi_{\pm}^{\nu}  = \sum_{i_{1} < i_{2} < ... < i_{N} }
D^{\nu}_{\pm} (i_{1} , i_{2} , ..., i_{N} ) \prod_{k=1}^{N}
\gamma^{ \dagger }_{i_k,\pm} \mid 0 \rangle,
\label{skfunc}
\end{eqnarray}
where $\gamma^{\dagger}_{m,\pm}$ are required to satisfy the relation
$\gamma^{\dagger}_{m,\pm} = u_{m} c_{m,\downarrow}^{\dagger}
+ v_{m} c_{m\mp 1,\uparrow}^{\dagger}$ and variational parameters
$u_m$ and $v_m$ under the condition $|u_{m}|^{2} +|v_{m}|^{2}=1$
determine the orientation of spins.
Then, $\Psi_{\pm}^{\nu}$ are invariant under rotation by
$R_{\theta} \equiv \exp [i\theta (L_Z \pm S_Z )]$, where  
$L_Z$ and $S_Z$ denote the $z$-components of total orbital and spin
angular momentums, respectively.
The rotational invariance under $R_{\theta}$ is not associated with
the exact symmetry of the Hamiltonian, but originated from the spin-charge
relation in the lowest Landau level \cite{moon}.
The changes of the charge densities $\rho ({\bf r})$ associated with
$\Psi_{\pm}^{\nu}$ in the core region are exactly related to the topological
charges $Q = \pm 1$ via $ \int d^2 {\bf r} ( \rho ({\bf r}) - \nu
/ (2\pi l^2) ) = \nu Q$,
where $l$ is the magnetic length $\sqrt{ \hbar c/(e B)}$.
Thus, we see that $\Psi_{-}^{\nu}$ ($\Psi_{+}^{\nu}$) represents
the wave function for a skyrmion (antiskyrmion).
Our conjecture that the skyrmion and antiskyrmion states are eigenstates
of $L_Z \pm S_Z$ is consistent with the result of the $\sigma$ model
\cite{rajarman} and was verified by recent numerical calculations 
around $\nu$ = 1 \cite{xie2}.

The coefficients $D^{\nu,\pm}$ can be well approximated by $C^{\nu}$
in Eq. (\ref{ground}) because their values must be equivalent to those
of $C^{\nu}$ in the case of $\{u_{m}=1; m = 0, 1, 2,...\}$. 
Moreover, considering a finite size of skyrmion, the electron correlation
effects in the ground state and skyrmion excitation state will be nearly
equal at suffciently large distances from the origin.
In fact, our choice of $C^{\nu}$ for $D^{\nu,\pm}$ gives the charge
and spin density profiles of skyrmion and antiskyrmion similar to those
obtained from the field theory \cite{ahn1,sondhi}. 
For $\nu$ = 1, our wave functions are found to be consistent with
those proposed by Fertig {\it et al.} \cite{fertig}.
It is also noted that $\Psi_{-}^{\nu}$($u_{m}=1; m = 0, 1, 2,...$)
is equivalent to the quasi-hole state of MacDonald and Girvin \cite{macdonald}.
For the antiskyrmion state, we need to set $u_0$ = -1 and $v_0$ = 0
to project the single particle state with $m$ = -1 onto the lowest
Landau level.

For the wave functions $\Psi^{\nu}_{\pm}$, the expectation values of the 
Hamiltonian $H$ can be calculated analytically, using the following relations:
\begin{eqnarray}
\nonumber
&&\langle \Psi_{\pm}^{\nu} |
c^{\dagger}_{m,\downarrow} c_{m,\downarrow}
| \Psi_{\pm}^{\nu} \rangle =
|u_m |^2 \langle \Phi_{g}^{\nu} | c^{\dagger}_{m,\uparrow} c_{m,\uparrow}
| \Phi_{g}^{\nu} \rangle 
\\  
\nonumber
&&\langle \Psi_{\pm}^{\nu} |
c^{\dagger}_{m,\uparrow} c_{m,\uparrow}
| \Psi_{\pm}^{\nu} \rangle =
|v_{m \pm 1} |^2 \langle \Phi_{g}^{\nu} | c^{\dagger}_{m\pm 1,\uparrow} 
c_{m\pm 1,\uparrow} | \Phi_{g}^{\nu} \rangle 
\\
&& \langle \Psi^{\nu}_{\pm} | \gamma_{m_1,\pm}^{\dagger}
\gamma_{m_2,\pm}^{\dagger} \gamma_{m_4,\pm} \gamma_{m_3,\pm}
| \Psi^{\nu}_{\pm} \rangle =
 \langle \Phi_{g}^{\nu} | c_{m_1,\uparrow}^{\dagger}
c_{m_2,\uparrow}^{\dagger} c_{m_4,\uparrow} c_{m_3,\uparrow}
| \Phi_{g}^{\nu} \rangle.
\label{expectation}
\end{eqnarray}
At $\nu$ = 1/$M$, the expectation values in Eq. (\ref{expectation}) are
given in literatures \cite{macdonald,analytic}, and their analytical forms
can be extended to the case of $\nu$ = 1-1/$M$ by using the particle-hole
symmetry between $\nu$ and $1-\nu$. 
Defining $\Psi_{0-}^{\nu}=\Psi_{-}^{\nu}(u_m=0; m=0,1,2,...)$ and 
$\Psi_{0+}^{\nu}=\Psi_{+}^{\nu}(u_0=-1,u_m=0;m=1,2,...)$ for the
skyrmion and antiskyrmion excitations, respectively, we calculate the
energy differences $\delta \epsilon^{\nu}_{\pm} \equiv < \Psi^{\nu}_{\pm}
| H | \Psi^{\nu}_{\pm} > - < \Psi^{\nu}_{0\pm} | H | \Psi^{\nu}_{0\pm} >$.
At $\nu$ = 1, the energy functionals of skyrmion and antiskyrmion are
equal because of the particle-hole symmetry between $\nu$ and $2-\nu$,
as previously noted by Fertig {\it et al.} \cite{fertig}.
In this case, we may consider a spin configuration of 
$\{u_{0}=u_{1}=...=u_{p}=1,u_m=0;m>p\}$, however, we find this state to
be energetically unstable due to large Coulomb interaction between opposite
spins on the boundary of the reversed spins.
This result leads us to consider skyrmions as symmetry breaking
states \cite{ahn1}.

\section{Results and discussion}

For $\nu$ = 1, 4/5, 2/3, 1/3, and 1/5, the calculated energy differences
$\delta \epsilon_{-}$ and $\delta \epsilon_{+}$ are plotted
in Fig. 1, as a function of the effective Zeeman coupling
$\tilde{g}$ [= $\frac{1}{2} {g^{*} \mu_{B} B}/(e^{2}/\epsilon l)]$.
We find that the skyrmion undergoes a transition into the spin-polarized
quasihole state $\Psi_{0 -}^{\nu}$ at critical value $\tilde{g}_c$,
while no such a transition exists for the antiskyrmion state.
Since the energy of the reversed electrons increases for stronger Zeeman
couplings, the skyrmion size tends to shrink to zero at $\tilde{g}_c$.
For $\tilde{g} > \tilde{g}_c$, all $u_m$'s for the skyrmion state
are found to be zero.
Thus, skyrmions exhibit the critical behavior with the order parameter
$u_m$ varying with $\tilde{g}$.
Based on the Landau theory, the number of reversed spins in the limit
$ \tilde{g} \rightarrow \tilde{g_c}^{-}$ is expressed such as
\begin{eqnarray}
\delta s_{-} = \nu   \sum_{m=0}^{m=\infty} |u_m|^2 
\sim  |\tilde{g}-\tilde{g}_c|^{\beta},
\end{eqnarray}
where $\beta$ represents the critical exponent.
From the calculated net spins $\delta s_{-}$, we estimate $\beta$
to be about 1 \cite{ahn1}.
If we assume that the skyrmion statistics is the same as that of the
Laughlin quasiparticles, as addressed before \cite{nayak},
the skyrmion condensation can be considered in the framework of
hierarchy construction.
Then, for the daughter states of the $\nu$ = 1/$M$ parent,
the filling factor $\nu^{\prime}$ is given by $\nu^{\prime} =
\frac{2p}{2 M p - \alpha}$, where $p$ is integer number, and
the spin polarization $P$ defined as $<S_Z> / (N/2)$ is written as
\begin{eqnarray}
 P = 1 - \frac{\delta s_{\pm}+ (\alpha+1)/(2M)}{p}, 
\label{condense1}
\end{eqnarray}
where $\alpha=1$ for antiskyrmion condensation and $\alpha=-1$ for
skyrmion condensation.
Thus, it is predicted that as the skyrmion-condensated quantum Hall
system at $\nu = \frac{2p}{2 M p + 1}$ transforms to a spin-polarized
state, its polarization behaves as $P - 1 \sim (\tilde{g} -\tilde{g}_c)
\sim (\sqrt{B} -\sqrt{B_{c}})$, where $B_{c}$ denotes the critical field.

\begin{figure}
\centerline{\psfig{figure=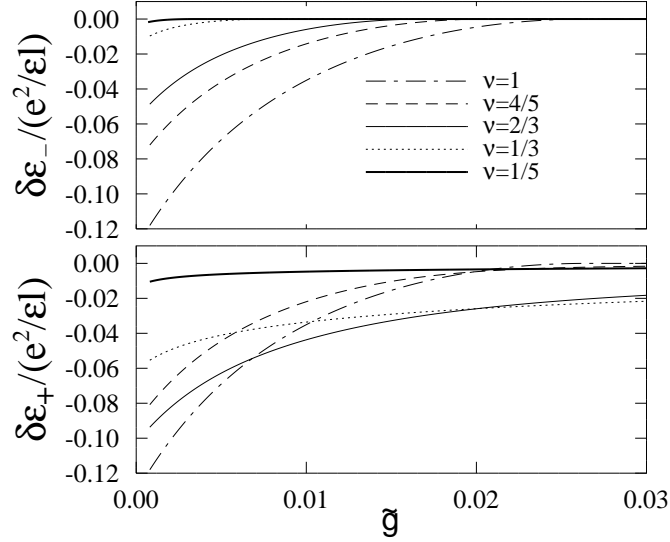,height=8cm}}
\caption{ Calculated $\delta \epsilon_{-}$ and $\delta \epsilon_{+}$
vs effective Zeeman coupling for various filling factors. }
\end{figure}

For filling factors near $\nu$ = 1/3 and 1/5,
the skyrmion excitations exist up to the critical fields of 7.2 and
1.3 T, respectively, for GaAs samples, while previous Monte Carlo 
calculations showed the corresponding values of 1.6 and 0.21 T.
Our calculations suggest that the skyrmions around $\nu$ = 2/3 and 4/5 are 
reliable quasiparticle states, if the ground states at these filling
factors are the particle-hole conjugate states of the Laughlin
$\nu$ = 1/3 and 1/5 states.
There are some experimental evidences for the skyrmion excitation at
$\nu$ = 2/3; the magnetoabsorption spectroscopy reveals a dramatic
reduction of the spin polarization near $\nu$ = 2/3 \cite{eisen}, and the
magnetotransport measurement of the activation energy indicates that relevant
quasiparticles are associated with several reversed spins \cite{aifer}.
It is pointed out that the skyrmion state around $\nu$ = 2/3 can not be
explained in the composite fermion picture \cite{jain}, where the $\nu$
= 2/3 state is obtained from the $\nu$ = 2 singlet state by composite
fermion transformation.
Using the calculated energies of the polarized and unpolarized ground
states at $\nu$ = 2/3 \cite{xie1}, we find that $\tilde{g}$ should be
greater than 0.0087.

\section{Conclusions}

We have presented the microscopic theory of skyrmion and antiskyrmion
in fractional quantum Hall systems and calculated in an analytical
fashion their excitation energies using variational wave functions.
We find that the single-Slater-determinant state for the skyrmion
near $\nu=1$ is energetically unstable because of the large Coulomb
interaction between neighboring electrons with opposite spins.
The skyrmion-condensated states are predicted to show the critical
behavior in the vicinity of the transition into the spin-polarized state. 

{\it Acknowledgments.}
This work was supported by the CTPC and CMS at KAIST.


\begin{thebibliography}{0}
\bibitem {sondhi}
D.-H. Lee and C. L. Kane, Phys. Rev. Lett. {\bf 64}, 1313 (1990);
S. L. Sondhi, A. Karlhede, and S. A. Kivelson,
 Phys. Rev. B {\bf 47}, 16419 (1993). 
\bibitem {barret} S. E. Barret {\it et al.},
Phys. Rev. Lett. {\bf 74}, 5112 (1995).
\bibitem {schmeller} A. Schmeller {\it et al.},
Phys. Rev. Lett. {\bf 75}, 4290 (1995). 
\bibitem {aifer}  E. H. Aifer {\it et al.},
 Phys. Rev. Lett. {\bf 76}, 680 (1996).
\bibitem {fertig} H. A. Fertig, L. Brey, R. Cote, and A. H. Macdonald,
 Phys. Rev. B {\bf 50}, 11018 (1994).
\bibitem {nayak} C. Nayak and F. Wilczek, 
LANL Report No. cond-mat/9512061, unpublished.
\bibitem{moon} K. Moon {\it et al.}, Phys. Rev. B {\bf 51}, 5138 (1995).
\bibitem{rajarman} R. Rajaraman, {\it Solitons and Instantons}
(North-Holland, Amsterdam, 1982).
\bibitem{xie2} X. C. Xie and S. He, Phys. Rev. B {\bf 53}, 1046 (1996).
\bibitem{ahn1} K.-H. Ahn and K. J. Chang, unpublished.
\bibitem{macdonald} A. H. MacDonald and S. M. Girvin, Phys. Rev. B {\bf 34},
 5639, (1986).
\bibitem{analytic} S. M. Girvin, Phys. Rev. B {\bf 30}, 558 (1984).
\bibitem{eisen} J. P. Eisenstein {\it et al.}, Phys. Rev. B {\bf 41},
    7910 (1990).
\bibitem{jain} J. K. Jain, Phys. Rev. Lett. {\bf 63}, 199 (1989).
\bibitem{xie1} X. C. Xie, Y. Guo, and F. C. Zhang, Phys. Rev. B
{\bf 40}, 3487 (1989).
  \end{thebibliography}
\end{document}